# On the Properties of Monocyclic Systems, and others related to them.
## (Crelles Journal 98. S. 68-94. 1884 u. 1885.)
## Ludwig Boltzmann

*Translated 2019 by Kim A. Sharp and Franz Matschinsky[1]*
*University of Pennsylvania*

The complete mechanical proof of the second law would obviously be provided if any mechanical process could be mathematically defined in a manner consistent with the equations of the theory of heat. However, the statement does not seem to be correct in this general form and the mechanical condition associated with thermal motion cannot be precisely defined because we do not understand the nature of the so called atoms. It is thus necessary to investigate in what cases and to what extent the equations of mechanics are analogous to those of heat theory. This will not require formulating mechanical systems which match precisely hot bodies, but by finding all systems which exhibit behavior more or less analogous to warm bodies. Herr von Helmholtz was the first to approach the issue in this manner[2] In the following I intend to pursue further the analogy which according to him exists between the systems which he called monocyclic and the laws of mechanical theory of heat, by using several systems which are closely related to monocyclic systems. In this endeavor I will discuss several special examples before I turn my attention to general laws.[3]

§1.

Let us assume a point mass is moving in an elliptical orbit around a fixed central body $O$ according to Newton's laws of gravitation. The movement in this case is obviously not monocyclic. But we are able to convert it to a monocyclic one by using a stratagem which I applied previously in the first paragraph of my treatise "Einige allgemeine Sätze über Wärmegleichgewicht"[4] ("Some general considerations about the equilibrium of heat") and which were subsequently pursued by Maxwell.[5]

We imagine that the entire elliptical orbit is occupied by mass, which is said to have at each point such a density (defined as the mass per unit length of the path) that at each point this density remains unchanged while a continuous flow of mass occurs at every cross section. An example of the movement in question would be if a ring of Saturn were composed of a homogeneous liquid or a homogeneous swarm of solid bodies, with a suitable choice of the ring cross-section at various points. External forces can accelerate the movement or change the eccentricity of the orbit. But the orbit can also be changed by the a slow increase or decrease in the mass of the central body. In this case work is done on moving the ring, as the ring exerts on the original mass of the central body a different attraction with the progressive addition or removal in time of the additional mass. The corresponding equations for this very simple example are obtained from those in my paper: "Remarks on some problems of the mechanical theory of heat"[6] Section 3 by setting $b = 0$ and by using $m$ to denote the total mass of the mobile ring. (Incidentally the heat produced by external work was in that report mistakenly stated with an incorrect sign.) I will hereafter denote the total potential energy by $\Phi$, the total kinetic energy by $L$, and the work of changing the internal energy by $dQ$. In this approach I assume as Herr v. Helmholtz did that the external forces vary only infinitesimally from those required to maintain the current motion. Then if $a/r^2$ is the net attractive force on the mass of the ring at distance $r$ towards the central mass, and $C$ is an undetermined constant, then

$$\Phi = C - 2L, \quad dQ = -dL + 2L\frac{da}{a} = L\, d\log\frac{a^2}{L}; \tag{1}$$

because $L$ appears as an integrating denominator of $dQ$, the corresponding expression for the entropy $S$ is $\log(a^2/L)$, and the corresponding characteristic function is:

$$K = \Phi + L - LS = C - L - L\log\frac{a^2}{L}. \tag{2}$$

---

1  Translators' notes are placed at the end of the text.
2   Berl. Ber. 6. und 27. März 1884.
3   A very general example of mono-cyclic systems is given by resistance-free electrical currents: see Maxwell, Treatise on Electricity, Art. 579 & 580, where x and y correspond to v. Helmholtz's variables $p_a$ and $p_b$.
4   Wien. Ber. 83. 1871. (L. Boltzmann, Wiss. Abh. (Collected Works) Vol. I. Nr. 19.)
5   Cambridge Phil. Trans. 12. III. 1879. (Vgl. auch Wiedemanns Beiblätter 5. S. 403 1881. Wiss. Abh. II. Nr. 63.)
6   Wien. Ber. 75. Wiss. Abh. I. Nr. 39. Vgl. auch Clausius, Pogg. Ann. 142. S. 433; Math. Ann. von Clebsch 4- S. 232, 6.

Substituting

$$2L\frac{\sqrt{L}}{a} = q, \quad \frac{a}{\sqrt{L}} = s, \tag{3}$$

then for $dQ = q\,ds$, its characteristic function is

$$H = \Phi + L - qs = C - 3L \tag{4}$$

and one sees immediately that

$$\left(\frac{\delta K}{\delta a}\right)_L = \left(\frac{\delta H}{\delta a}\right)_q = -A \tag{5}$$

is the gravitational force acting to increase $a$, in the sense that $A\,da$ is the amount of work produced by the internal motion. Similarly

$$\left(\frac{\delta K}{\delta L}\right)_a = -S, \quad \left(\frac{\delta H}{\delta q}\right)_a = -s. \tag{6}$$

The analogy with Herr v. Helmholtz's mono-cyclic systems with a single velocity $q$ is thus complete. Also $\int q\,dt$ has the characteristic of a coordinate; because

$$i = \pi\frac{\sqrt{2m}}{q} \tag{7}$$

is the orbital period of a particular mass (opus cit.). Let $\rho\delta\sigma$ be the mass along the length of an orbit element $\delta\sigma$, $dt$ is the time required to traverse $\delta\sigma$, and $v$ be the velocity there. Then $\rho v = c$ must be a

---

S. 390. Nachricht. d. Gött. Gesellsch. Jahrg. 1871 u. 1872; Pogg. Ann. 150. S. 106 u. Ergünzungsbd. 7. S. 215

constant around the orbit so that the density is unchanged everywhere.

Since $\int \rho v \, dt$ taken over the entire orbit is the total mass $m$ of the ring, then

$$c = \frac{m}{i} = q \frac{\sqrt{2m}}{2\pi}. \tag{8}$$

If we extend from the central body an arbitrary straight line extending to Infinity, and denote by $\mu$ the mass which up to a particular moment of time $t$ passed through that straight line then

$$\frac{d\mu}{dt} = \rho v, \quad \text{hence} \quad q = \frac{2\pi}{\sqrt{2m}} \frac{d\mu}{dt}. \tag{9}$$

Let us now consider some particular mass element within the ring, so that once the values of $a$ and $\mu$ are known, along with the initial position of this mass, the location of this mass after time $t$ is determined, without needing to know how entire orbital path has changed in the meantime. One can therefore consider $a$ and $\mu$ as the coordinates of the respective mass element. The same is true for any other type of central force dynamics, just as long as the orbit is a closed one. If for example each mass element $v$ is pulled with a force $var/m$ towards the central body, then

$$\Phi = L, \quad dQ = 2dL - L\frac{da}{a} = Ld\log\frac{L^2}{a}, \quad q = 2\sqrt{a}, \quad s = \frac{L}{\sqrt{a}}. \tag{10}$$

§2.

The use of the term mono-cyclic for the motions described here is probably justified since all the conditions specified by Herr v. Helmholtz are met. The fact that the kinetic energy is not proportional to $q^2$ when the parameter $a$ is eliminated is probably due to the fact that $a$ is not a spatial quantity in the true sense of the term. Analogous formulas will be valid in the more general case. The forces need not be central forces, they may instead be represented by arbitrary functions of the coordinates of the particular mass elements on which they act. Still, a force function must exist, the force must be the same function of the coordinates for all particles of mass $m$, and all particles must follow congruent closed paths.

Constructing a plane of arbitrary but fixed direction through the center of gravity of the total mass $m$, and denoting by $\mu$ the mass that has passed through the plane during any time $t$, one has

$$p_b = \mu, \quad q_b = \frac{d\mu}{dt}. \tag{11}$$

The condition for motion in a closed path is always satisfied if all mass particles move in parallel paths (it is immaterial whether in a single path or multiple parallel paths). Let then $\mu$ be the mass passing, during any time $t$, through a plane through the center of gravity and perpendicular to the orbit, then again

$$p_b = \mu, \quad q_b = \frac{d\mu}{dt} = \frac{m}{i}, \tag{12}$$

where *i* is the time required for all the mass elements to leave the plane and return again. Let us focus on a single point mass from the entire mass which, at the beginning of the time period is at the above-defined plane. If one knows at any time *t* the value of $p_b$, and the force laws, one can determine where the selected point is at any time, without knowing the intermediate states. At time *t*, let *x* be the distance of the particle from the center of the total mass (with respect to this particular plane). Let *v* be its velocity and *f'(r)* be the force per unit mass acting to decrease *x*, then

$$v^2 = 2a - 2f(x), \quad i = 2\int_{x_1}^{x_2} \frac{dx}{\sqrt{2a-2f}}, \tag{13}$$

where $x_1$ and $x_2$ are the extremal values of *x*. Then the kinetic energy of the entire mass is:

$$L = \frac{m}{i}\int_{x_1}^{x_2} \sqrt{2a-2f}\, dx. \tag{14}$$

The potential energy is

$$\Phi = \frac{2m}{i}\int_{x_1}^{x_2} \frac{f\, dx}{\sqrt{2a-2f}}; \tag{15}$$

where

$$f = f(x), \tag{16}$$

*a* is a constant.

Then it follows:

$$H = \Phi - L = ma - 2L. \tag{17}$$

Here also *a* and the function *f* vary in magnitude slowly. Taking the latter as constant, and writing the above equation in the form

$$iH = mia - 2m\int_{x_1}^{x_2} \sqrt{2a-2f}\, dx, \tag{18}$$

the differential is:

$$H + i\frac{\partial H}{\partial i} = ma + mi\frac{\partial a}{\partial i} - 2m\frac{\partial a}{\partial i}\int_{x_1}^{x_2} \frac{dx}{\sqrt{2a-2f}} \tag{19}$$

or, with regard to the expression found for $i$:

$$H + i \frac{\partial H}{\partial i} = ma; \tag{20}$$

Herr v. Helmholtz introduced the variable $q = mi$, which results in

$$s = \frac{-\partial H}{\partial q} = \frac{ma - H}{q} = \frac{2L}{q} = \frac{2iL}{m}. \tag{21}$$

From this it finally follows that:

$$dQ = qds = \frac{2}{i} d(iL) = 2L \cdot d\log(iL). \tag{22}$$

This is the same result that Clausius obtained with a totally different approach for a somewhat more special case (compare his treatise in Berl. Ber. of 19th June 1884.) If one chooses the coordinates consistently following the approach of Herr v. Helmholtz, his formula is perfectly applicable to this case. To facilitate the extension to other classes of systems, I will defer the general treatment, and consider the following special case of the preceding example.

A homogeneous stream of constant density is moving at velocity $v$ along a horizontal straight line of length $a$ between two vertical walls, so that one half the mass, $m/2$ is always moving in the opposite direction to the other. The mass particles are not supposed to act on each other nor are they to be affected by external forces except for those which perform the work $dQ$. Collisions with the vertical walls are perfectly elastic. This system is strictly mono-cyclic. One has:

$$q = \frac{mv}{2a}, \quad H = -\frac{2a^2 q^2}{m}, \quad dQ = mvdv + mv^2 \frac{da}{a} = qds. \tag{23}$$

whereby

$$s = -\left(\frac{\partial H}{\partial q}\right)_a; \tag{24}$$

$-(\partial H/\partial a)_q$ is the pressure exerted on the vertical wall.

We modify the above example such that a mass $m$ is uniformly distributed in a parallelepiped vessel with sides $a, b, c$. Every mass particle has to move in a straight line with the velocity $v$, without being influenced by internal or external forces (again with the exception of that which generates the work $dQ$). This straight line is perpendicular to the side $c$. For half of the particles this line forms an angle $+D$ with the side $a$ and for the other half it forms an angle $-D$. If we consider $a, b$, and $v$ as variables, then:

$$\Phi = 0, \quad L = -H = \frac{mv^2}{2}, \tag{25}$$

$$dQ = mv\,dv + mv^2\left(\sin^2 D \frac{da}{a} + \cos^2 D \frac{db}{b}\right); \tag{26}$$

These equations take the same form as those of von Helmholtz if one chooses $a$ and $b$ as the parameter $p_a$, and sets

$$q = \frac{v}{a^{\sin^2 D} b^{\cos^2 D}}. \tag{27}$$

The pressures in the directions $a$ and $b$ are

$$-\left(\frac{\partial H}{\partial a}\right)_{bq} \quad \text{and} \quad -\left(\frac{\partial H}{\partial b}\right)_{aq} \tag{28}$$

$dQ$ is equal to $qds$, where

$$s = -\left(\frac{\partial H}{\partial q}\right)_{ab}. \tag{29}$$

The kinetic energy would, however, no longer be the integrating denominator of $dQ$, even if one allowed those external forces which slowly and uniformly change the angle $D$ for all point masses.

If the mass $m$ were to flow with constant velocity $v$ in all possible directions inside a vessel of volume $w$ one would have to set

$$q = \frac{v}{w^{1/3}}, \quad p_a = w, \tag{30}$$

and the equations would again adopt the form of von Helmholtz's. However in all these cases, $\int q\,dt$ no longer has the character of a coordinate since the trajectory does not return to itself in a finite time. I want to define a system whose motion in this sense is stationary as monodic, or more concisely, a Monode.[7] This has the property that the equations of motion are the same for every point and invariant with time as long as the external forces are unchanged; also that at no point on the surface of the region does kinetic energy or mass enter or exit. If the kinetic energy functions as the integrating denominator of the differential $dQ$, and the work acts as to alter the internal motion, I will designate this an Orthode. All the Orthodic equations are completely analogous to those of thermodynamics.

Let $dQ$ be the differential due to either an increase in internal work of an Orthode, or from heat added to a warm body. We postulate that $L$ is an integrating denominator for $dQ$; This applies to warm bodies when the temperature is proportional to the kinetic energy. Let $p$ be any parameter. We set

---

7   The term stationary was used by Herr Clausius for motions whose coordinates and velocities remained bounded between finite limits.

$$dQ = d\Phi + dL + \sum Pdp = 2Ld\log s. \tag{31}$$

If $q$ is any other integrating denominator of $dQ$ and $dQ = qdS$, then according to Gibbs[8] $\Phi + L - qS$ also possesses the properties characteristic of a Massieu function. Let $q = 2L/s$, which results in

$$dQ = qds, \tag{32}$$

and the corresponding characteristic function is $H = \Phi - L$, exactly as found by von Helmholtz for the monocyclic systems (compare 1.c. S. 173.) However, now $\int qdt$ is usually no longer a coordinate. For $q = 2L/s^n$ becomes $H = \Phi + L(n-2)/n$, and if $n=2$ then $H = \Phi$.

I mention here in passing that the above equation of Gibbs' may be inapplicable in particular cases, specifically when $q$ is a function of the parameter $p$ alone. It is then not possible to introduce $q$ as an independent variable in addition to $p$. Thus for gases, according to von Helmholtz (*loc. cit.* S. 170):

$$dQ = J\gamma d\theta + J(c-\gamma)\frac{\theta dv}{v} = \frac{J\gamma}{v^{\frac{c-\gamma}{\gamma}}} d\left(\theta v^{\frac{c-\gamma}{\gamma}}\right). \tag{33}$$

If

$$q = \frac{J\gamma}{v^{\frac{c-\gamma}{\gamma}}}, \quad S = \theta v^{\frac{c-\gamma}{\gamma}}, \tag{34}$$

then

$$H = J\gamma\theta - qS = 0, \tag{35}$$

must be rejected as a characteristic function. The same situation was found in the example presented at the end of §1 with $q$ as the designated integrating denominator.

§3

Following these introductory examples, I want to move on to a very general case. Let's imagine an arbitrary system whose state is specified by arbitrary coordinates $p_1, p_2, \ldots p_g$, and corresponding momenta $r_1, r_2, \ldots r_p$. For brevity we will call the coordinates $p_g$, and the momenta $r_g$. Let us assume the system is subject to any desired internal and external forces; the former should be conservative. Let $\psi$ be the kinetic energy and $\chi$ be the potential energy of the system. Then $\chi$ is a function of $p_g$, and $\psi$ is a second degree homogeneous function of $r_g$, whose coefficients may also contain $p_g$. We want to determine the arbitrary constant added to $\chi$ so that for infinite separation of all mass elements, $\chi$ disappears and is independent of their positions. We avoid the restrictive assumption that certain coordinates of the system are forced to adopt defined values, because these could not be determined if their values depended on an external force which was being slowly varied. Instead, the slow variability of the external forces is to be accounted by an additional gradual functional dependence of $\chi$ on $p_g$; Or that certain constants contained in $\chi$, say $p_a$, change slowly.

---

8     Transact. of the Conn. Acad. III, S. 108. 1875.

Case 1. We imagine now very many *(N)* identically constructed systems; Each system is totally independent of all the others. The number of all these systems for which the coordinates and momenta lie between the limits

$$p_1 \text{ and } p_1 + dp_1, \; p_2 \text{ and } p_2 + dp_2, \ldots r_g \text{ and } r_g + dr_g \tag{39}$$

is

$$dN = N e^{-h(\chi+\psi)} \frac{\sqrt{\Delta} \, d\sigma \, dr}{\int \int e^{-h(\chi+\psi)} \sqrt{\Delta} \, d\sigma \, dr}, \tag{40}$$

where

$$d\sigma = \Delta^{-1/2} dp_1 dp_2 \ldots dp_g, \quad dr = dr_1 dr_2 \ldots dr_g \tag{41}$$

(For the definition of $\Delta$ see Maxwell, *loc. cit.* p556)

The integrals are taken over all possible values of the coordinates and momenta. The collection of all these systems forms a Monode in the previously defined sense (see here especially Maxwell, *loc. cit.*). I wish to call this class of Monodes Holodes. Each of the systems forms an element of the Holode. The total kinetic energy of the Holode is:

$$L = \frac{Ng}{2h} \tag{42}$$

The potential energy $\Phi$ is N times $\overline{\chi}$, the average of $\chi$ and is therefore:

$$\Phi = N \frac{\int \chi e^{-h\chi} d\sigma}{\int e^{-h\chi} d\sigma}. \tag{43}$$

The coordinates $p_g$ differ from Helmholtz's $p_b$ in that they appear in the expressions for kinetic energy $\psi$ and the potential energy $\chi$. The intensity of the motion of the entire Ergode, that is both $L$ and $\Phi$, depends only on $h$ and $p_a$, in the manner of Helmholtz's $q_b$ and $p_a$.

The relationship between the work and the increase in internal motion is

$$\delta Q = \delta \Phi + \delta L - N \frac{\int \delta \chi e^{-h\chi} d\sigma}{\int e^{-h\chi} d\sigma} \tag{44}$$

(Compare this with my treatise "Analytic proof of the second law of thermodynamics from the balance of the kinetic energy"[9] and Maxwell *loc. cit.*). The amount of internal motion which is produced by the external work when the parameter $p_a$ changes by $\delta p_a$ is $-P\delta p_a$ whereby

$$-P = \frac{\int \frac{\partial \chi}{\partial p_a} e^{-h\chi} d\sigma}{\int e^{-h\chi} d\sigma}. \tag{45}$$

The kinetic energy $L$ is the integrating denominator of $\delta Q$; All Holodes are therefore Orthodic, and because of this the remaining heat theory analogies have to exist also. Indeed, if one sets

$$s = \frac{1}{\sqrt{h}}\left(\int e^{-h\chi} d\sigma\right)^{\frac{1}{g}} e^{\frac{h\chi}{g}} = \sqrt{\frac{2L}{Ng}}\left(\int e^{-h\chi} d\sigma\right)^{\frac{1}{g}} e^{\frac{\Phi}{2L}}, \tag{46}$$

$$q = \frac{2L}{s}, \quad K = \Phi + L - 2L\log s, \quad H = \Phi - L, \tag{47}$$

it then follows that

$$dQ = 2L\, d\log s = q\, ds, \quad \left(\frac{\partial K}{\partial p_a}\right)_h = \left(\frac{\partial H}{\partial p_a}\right)_q = -P, \tag{48}$$

$$\left(\frac{\partial K}{\partial h}\right) = -2\log s, \quad \left(\frac{\partial K}{\partial q}\right) = -s. \tag{49}$$

Case 2. As described in the beginning of this section, we assume again the existence of very many ($N$) systems of the same type. However, for all of these the conditions

$$\phi_1 = a_1, \, \phi_2 = a_2, \, \ldots \phi_k = a_k \tag{50}$$

apply. These equations in each case need to be integrals of the equation of motion of the system. Other integrals may exist in addition. Now $dN$ is the number of systems for which the coordinates and momenta lie between the limits $p_1$ and $p_1+dp$, $p_2$ and $p_2+dp$, … $r_g$ and $r_g+dr_g$. Naturally this does not include the differentials of those coordinates or momenta which are determined by the equations $\varphi_1=a_1$…. These $h$ missing coordinates or momenta are denoted by $p_c$, $p_d$, … $r_f$. If then

$$dN = \frac{\dfrac{N\, dp_1 dp_2 \ldots dr_g}{\sum \pm \dfrac{\partial \varphi_1}{\partial p_c} \cdot \dfrac{\partial \varphi_2}{\partial p_d} \cdots \dfrac{\partial \varphi_k}{\partial r_f}}}{\int\int \dfrac{dp_1 dp_2 \ldots dr_g}{\sum \pm \dfrac{\partial \varphi_1}{\partial p_c} \cdot \dfrac{\partial \varphi_2}{\partial p_d} \cdots \dfrac{\partial \varphi_k}{\partial r_f}}} \tag{51}$$

---

9 Wien. Ber. 63. 1871, Eqn. (17). (Wiss. Abh. I. Nr. 20.)

we will call the collection of all N systems which are constrained by equation $\varphi_1 = a_1$ a Monode. The magnitudes $a$ are either constant or undergo a slow change. The functions $\varphi$ in general vary slowly as determined by the $p_a$. Each single system will be called an element. We will call those Monodes which have only the equation of motion as a constraint Ergodes. Those Ergodes which are constrained by other equations are called sub-Ergodes. Ergodes are defined by

$$dN = \frac{N dp_1 dp_2 ... dp_g dr_1 ... dr_{g-1}}{\frac{\partial \varphi}{\partial r_g}} \Big/ \int \int ... \frac{dp_1 dp_2 ... dr_{g-1}}{\frac{\partial \varphi}{\partial r_g}} \tag{52}$$

So for an Ergode there is only one value of $\varphi$, which is the same for all systems, and which during the evolution of each system is constant and equal to its energy $\chi + \psi = (\Phi + L)/N$. Again setting $\Delta^{-1/2} dp_1 dp_2 ... dp_g = d\sigma$, it follows then (see the previously cited treatises of Maxwell and myself):

$$\Phi = N \frac{\int \chi \psi^{g/2-1} d\sigma}{\int \psi^{g/2-1} d\sigma}, \quad L = N \frac{\int \psi^{g/2} d\sigma}{\int \psi^{g/2-1} d\sigma}, \tag{53}$$

$$\delta Q = N \frac{\int \delta \psi \, \psi^{g/2-1} d\sigma}{\int \psi^{g/2-1} d\sigma} = \delta(\Phi + L) - N \frac{\int \delta \chi \, \psi^{g/2-1} d\sigma}{\int \psi^{g/2-1} d\sigma}, \tag{54}$$

L is again the integrating factor of $\delta Q$, with an associated entropy of $\log \left( \int \psi^{g/2} d\sigma \right)^{2/g}$, where also $\delta Q = q \delta s$ when we set

$$s = \left( \int \psi^{g/2} d\sigma \right)^{2/g}, \quad q = \frac{2L}{s}. \tag{55}$$

With the latter Entropy we can associate the characteristic function $\Phi - L$. The external force in the direction of component $p_a$ in a system is

$$-P = \frac{\int \frac{\partial \chi}{\partial p_a} \psi^{g/2-1}}{\int \psi^{g/2-1}} \tag{56}$$

From the infinite number of possible sub-Ergodes, select all those systems which not only have the same kinetic energy, but also those for which the three surface constraint equations have the same, constant value. I wish to call these Planodes. Several of their characteristics were developed by Maxwell (loc. cit.). I note here that in general these are no longer Orthodic.

The state of a certain element of an Ergode is characterized by certain parameters $p_g$. When each element of the Ergode is an aggregate of many particles and when the number of parameters $p_g$ is smaller than the number of orthogonal coordinates specifying all these particle positions, there will always be some function of these coordinates which is an invariant of motion, and the preceding development assumes that these are invariant even when kinetic energy is added or removed. If one were to allow - besides the variability of the parameters in the energy function - an additional slow variation of the functions, which would then play the role of von Helmholtz's $p_a$, and which also are called here $p_a$ then one would arrive at equations which encompass my earlier treatment as well as that of von Helmholtz.

The equations which follow equation (18) of my paper "Analytic proof of the second law of thermodynamics from the equilibrium of the kinetic energy" were on that occasion not developed in their full generality. First, only one system is considered, which itself passes through all possible states consistent with its kinetic energy; Second only regular orthogonal coordinates were used. However, it is immediately clear from what Maxwell said in his treatise "On Boltzmann's theorem, etc.", which has been cited here several times, that these formulas must apply to any Ergode which is represented by generalized coordinate. For any element of an Ergode represented by $p_1, p_2, ...p_g$, one arrives at

$$\frac{d\mathfrak{R}}{N} = \frac{\Delta^{-1/2} \psi^{g/2-1} dp_1...dp_g}{\int\int...\Delta^{-1/2}\psi^{g/2-1}dp_1...dp_g}, \qquad (57)$$

where $N$ is the total number of systems in the Ergode, $d\mathfrak{R}$ is the number of systems with coordinates between $p_1$ and $p_1+dp_1$, $p_2$ and $p_2+dp_2$, $p_g$ and $p_g+dp_g$. $\Psi$ is the expression for kinetic energy in terms of the velocity. The ninth equation after formula (18) in the cited passage is

$$\frac{2}{3}\log \int \psi^{3\lambda/2} d\sigma + const. \qquad (58)$$

Because the variables $\lambda$ and $T$ have the values $g/3$ and $3\Psi/g$, and $\delta Q$ is the differential heat supplied to a single system, this leads to

$$\delta Q = \frac{2L}{g}.\delta \log \int\int...\Delta^{-1/2}\psi^{g/2}dp_1...dp_g \qquad (59)$$

$$= \frac{2N}{g}\frac{\delta\int\int...\Delta^{-1/2}\psi^{g/2}dp_1...dp_g}{\int\int...\Delta^{-1/2}\psi^{g/2-1}dp_1...dp_g},$$

$\delta Q$ is the heat supplied to the entire Ergode. $L = N\Psi$ is the total kinetic energy of the Ergode. Here it is immaterial whether one thinks of the potential energy function as directly varying, or whether one assumes that all coordinates $p_a$ other than $p_g$ are constant for a steady state Ergode and that a slow change is effected through the dependence of the potential function on $p_g$. Of course this is generally associated with a change in the value of the potential function with $p_g$. To understand this, it is a good idea to think of the $p_g$ chosen so that the limits are not directly affected by the change in $p_a$, but rather by changes of the energy function associated with it. The detailed proof will be given in a later essay. The motion itself is completely independent of which coordinates are used. The single velocity monocyclic systems of von Helmholtz are nothing more than Ergodes with a single rapidly changing coordinate, namely $p_g$, They are not like von Helmholtz's $p_a$ whose derivatives vary slowly with respect to time. Therefore the above formulas apply equally well to mono-cyclic systems at a single velocity and for warm bodies; And therefore it seems that in this way the analogy between rotational motion and ideal gases discovered by von Helmholtz is explained (compare Crelle's Journal, v97 p123, Berl. Ber. p170.)  If one system exists whose rapidly changing variables pass through all combinations of values which are compatible with the equation for kinetic energy (iso-monodic) then $N = 1$, $\Psi = L$. For a rotating solid body $g = 1$; We set $p$ equal to the angle coordinate $\theta$, and $\omega = d\theta/dt$, then

$$\psi = L = \frac{T\omega^2}{2} = \frac{r^2}{2T}, \quad r = T\omega; \tag{60}$$

therefore $\Delta = 1/T$, and $\Delta = \mu_1 \mu_2 \mu_3...$, then $L$ has the form

$$\frac{\mu_1 r_1^2 + \mu_2 r_2^2 + ...}{2} \tag{61}$$

$T$ is the moment of inertia; $\int\int ...dp_1 dp_2...$ reduces to $\int dp = 2\pi$ which can be omitted, so that the general formula above reduces to $\delta Q = L\delta \log(TL)$. If a single mass $m$ is rotating a distance $\rho$ from the axis, one can set $p$ equal to the moment $s$ of the mass. Then

$$\psi = L = \frac{mv^2}{2} = \frac{r^2}{2m}, \quad r = mv, \quad \Delta = \frac{1}{m}, \tag{62}$$

$$\int\int ...dp_1 dp_2... = \int dp = 2\pi\rho,$$

where $v = ds/dt$; This give the general formula above, $\delta Q = L\delta \log(mL\rho^2)$. For an ideal mono-atomic gas again $N = 1$, $\Psi = L$; $p_1 p_2...p_g$ are the Cartesian coordinates $x_1 y_1...z_n$ of the molecules; Therefore if $n$ is the total number of gas molecules, $g = 3n$, $v$ is the volume of the gas. $\int\int ...dp_1 dp_2...$ has the value $v^n$, $\Delta$ is constant as long as the masses of the molecules are constant; This gives the general formula above $\delta Q = L\delta \log(L.v^{2/3})$, which again is exactly the right form since in this case the ratio of the heat capacities is 5/3. If a mass M of liquid of density $\rho$ flows in a closed-loop channel of varying cross sections $\omega$, then two situations are possible.

    1. Each individual element of liquid is an Ergode system. We set $p = x$, equal to the path of the liquid element, whose mass is $\mu$, then $g = 1$, $N$ equals the total number of liquid elements,

$$\psi = \frac{\mu u^2}{2}, \quad \Delta = \frac{1}{\mu}, \quad \rho\omega u = q = const., \tag{63}$$

and since $\mu$ is constant the general formula yields

$$\delta Q = 2L\delta \log \int u\,dx = \int \rho\omega u^2 dx \frac{\delta \int u\,dx}{\int u\,dx} = q\delta \int u\,dx. \tag{64}$$

    2. The entire mass of fluid is a system of Ergodes, which belong to the same (isodic) system. Then $p$ must be chosen so that $dp/dt$ does not change rapidly. Following Herr von Helmholtz, set

$$\frac{dp}{dt} = q = \rho \omega u, \tag{65}$$

so then

$$L = \frac{q^2}{2} \int \frac{dx}{\rho \omega} = \frac{r^2}{2 \int \frac{dx}{\rho \omega}}, \quad r = \frac{\partial L}{\partial q} = q \int \frac{dx}{\rho \omega}, \tag{66}$$

$$\Delta = \frac{1}{\int \frac{dx}{\rho \omega}}, \quad \int dp = M, \quad \psi = L, \tag{66a}$$

and the general formula is recovered:

$$\delta Q = 2L \delta \log \sqrt{L \cdot \int \frac{dx}{\rho \omega}} \cdot M = 2L \delta \log q \int \frac{dx}{\rho \omega} = 2L \delta \log \int u \, dx. \tag{67}$$

Here, as in Helmholtz's paper, $\omega$ is the cross-section, $dx$ is a length element in the channel (streamline), and $u$ is the fluid velocity. For central motions we can define $p_g$ as a function of $\theta$, the polar angle with respect to a point along the orbit, then the equation of the orbit just described corresponds to $p_g(\rho\theta) = const$. The previous general formula for $\delta Q$ is recovered when $m$ is constant

$$\delta Q = \frac{m \int \rho^2 \omega \, d\theta}{\int \frac{d\theta}{\omega}} \delta \log \int \rho^2 \omega \, d\theta = \frac{m \int d\theta \, \delta(\rho^2 \omega)}{\int \frac{d\theta}{\omega}}, \tag{68}$$

wherein $\theta$ is not capable of variation; $\omega$ is equal to $d\theta/dt$; $m$ is the total mass of the Ergode. Incidentally, the general formula just given for $\delta Q$ also holds when $p_g$ is understood to refer to both planar and non-planar paths $s$. Then, setting $ds/dt = v$,

$$\delta Q = \frac{m}{\int \frac{ds}{v}} \cdot \delta \int v \, ds, \tag{69}$$

where the first factor is Hr. Helmholtz's $q_b$, the expression following the symbol $\delta$ is the magnitude denoted by Helmholtz as $s_b$. Here again we have the integral $\int v ds$, which appeared in my derivation starting from the principle of least action [10], which nicely demonstrates the connection among these various treatments. The variation of the integral, $\int \delta v ds$, directly yields the change in kinetic energy: The change $\int v \delta ds$, however, gives the amount of heat output as work. In fact if $R$ is the radius of curvature of the orbit, $\delta R$ is the displacement of the element $ds$ in the direction of $R$, then, with the appropriate type of variation, $\delta ds = \delta R ds/R$. The mass contained in $ds$ is

$$dm = \rho v\, dt = \frac{m\, dt}{i} = \frac{m\, ds}{v\, i}, \tag{70}$$

and the work done against the centripetal force is

$$\frac{v^2\, dm\, \delta R}{R} = \frac{m v\, ds\, \delta R}{i R} = \frac{m v\, \delta\, ds}{\int \frac{ds}{v}} \tag{71}$$

---

10  Wien. Ber. 53. 8. February 1866. (Wiss. Abh. I, Nr. 2.)

In the case where the mass moves Ergodically on a surface, $g=2$. So then a finite region of the surface is completely covered by a constant density of particles. The mass $dm$ which is located on a surface element $do$ is proportional to the quantity $do$ determined in Equation (24) of my paper "Some Theorems on Heat Equilibrium", i.e. equal to $mdo/\!\int\! do$. Furthermore

$$\delta Q = \frac{m}{2\int do} \delta \int v^2\, do, \tag{72}$$

where again $\int do\,\delta(v^2)$ is the heat used to increase the kinetic energy, $\int v^2\, \delta do$ is the heat converted into the work of overcoming the centripetal force. This is easily demonstrated by exact calculation of the centripetal force, bearing in mind that for an Ergode, every direction on the surface is equally probable. The general formula used here is of course also valid for composite mono-cyclic and poly-cyclic systems as long as they are ergodic. But I hesitate to increase the number of the special examples even further.

§4

The composite monocyclic systems are also Ergodes with one single rapidly changing quantity when there is a close coupling between the quantities $p_a$ and $p_a$ via the $n-1$ equations. Then only one single rapidly changing quantity is left and we have again one Ergode because this quantity runs through all possible values which are compatible with the equation of the kinetic energy and the coupling equations. This is because the parameters occurring in the constraint equations are expected to change slowly and can be regarded as constant relative to the more rapidly varying term. In all these cases, therefore, the kinetic energy must also be an integrating denominator, and there is complete agreement between my equation and those of von Helmholtz. This is still true if, among the coupling equations, linear equations with constant (not slowly changing) coefficients exist among the $q_b$ in a manner analogous to engaging cog wheels with a finite number of teeth. For fluid flows, such equations would correspond to paddle wheels whose speed of revolution is proportional only to the quantity of fluid passing through each cross-section. Such paddle wheels can be found in gas meters for measuring the amount of flammable gas consumed. If, on the other hand, the coefficients of the equations relating the $q_b$ are slowly changing, as in friction rollers, endless cords or water wheels, even if the energy consumed by frictional slippage is infinitesimal, then in general it is no longer possible to limit the final $q_b$ to a single value, or to a finite number of values given all the other $q_a$ and $q_b$, but the last $p$ would be capable of assuming an infinite continuum of values. Then the system loses its ergodic property, and the kinetic energy is generally no longer an integrating denominator; Indeed an integrating denominator of $dQ$ need longer exist, as I already noted in Anzeiger der Wiener Akademie of 9[th]. October 1884; In this case the variable coefficients of the constraint equations can either be functions of the afore-mentioned $q_a$, or new independent, slowly changing quantities can be added to describe the variation of the constraint conditions. It seems to me that without exception the original $q_a$ can be included.

This is the only case where I disagree with the deliberations of von Helmholtz, provided I actually do understand him correctly, when he states in Crelles Journal 97, p. 133: "As long as only those combinations (that is purely kinematic ones) are introduced, the kinetic energy remains one of the integrating denominators of the systems." Herr von Helmholtz arrives at this result (*loc. cit.* p. 125, see also p. 117) by assuming that the equation $dQ = 0$ has to have an integral of the form $\sigma = const.$ where $\sigma$ is a function of the independent variables $p_a$ and $x$ occurring in $dQ$. But this assumption does not always seem to me to be permissible; On the contrary, the expression of Herr von Helmholtz (*loc. cit.* 126b, Equation 6b) exactly specifies the conditions for $dQ$ to have any integrating factor at all; If the function $F$ is a homogeneous function of $s_b$, then among the integrating factors is the reciprocal of the kinetic energy. For by choosing an appropriate function of $\sigma$, the degree of the function $F$ can always be made equal to one. For the simplest case, where the constraint function $F$ is linear with constant coefficients, one can always assume that the latter have a common measure, however small, making the system ergodic. If on the other hand, $F$ is a more complicated homogeneous function, we obtain from the von Helmholtz equations Orthodic systems to which my equations no longer apply. Admittedly, all these systems are, as far as I can see, associated with mechanically highly unnatural conditions. Also not all parameters are likely to vary. On the other hand, my equations are not limited to the conditions where the derivatives of all rapidly changing quantities are functions of a single variable over time. Rather, in my equations, any number of independent derivatives $q_g$ of rapidly varying quantities $p_g$ can occur. The $q_g$ themselves can change rapidly; only the mean kinetic energy is determined by a single variable, namely the temperature. Special cases of my formulas correspond to the ideal gases and those classes of Monodes which in Maxwell's treatment (*loc. cit.*) represent solids and molten liquids occurring in Nature.

It is still necessary to demonstrate here with examples that this theoretically possible case, namely where the equation $dQ = 0$ does not contain any integrating factor, can be fully realized. In order to be as clear as possible I will describe here a particular, fittingly simple example which I constructed exactly as von Helmholtz suggested (Figure 1).

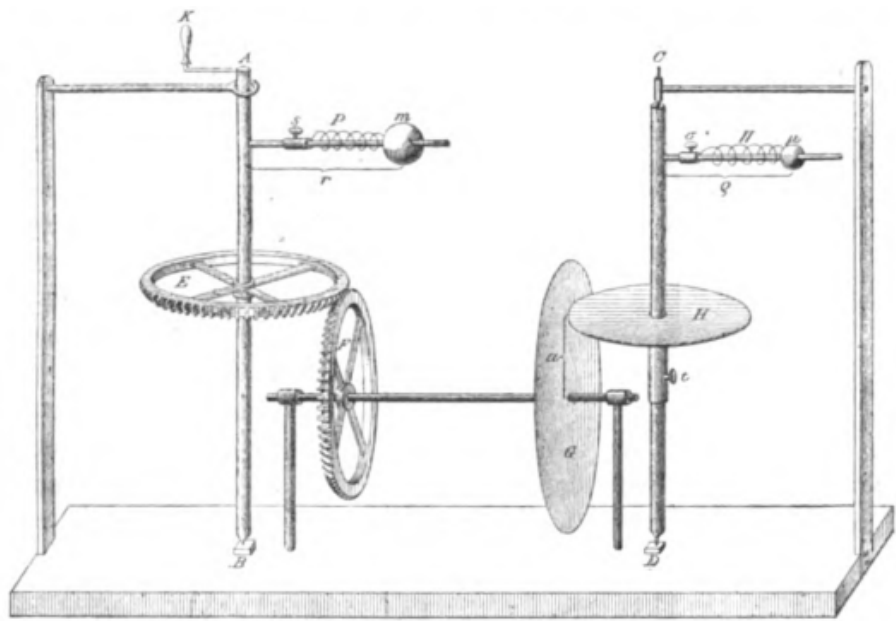

Fig. 1.

A fixed vertical axis *AB* bears a horizontal transverse arm on which a mass *m* can slide, just like the school-room device used to demonstrate centrifugal forces; Fixed to the mass *m* is an elastic spring *P* whose other end can be freely adjusted on the horizontal arm with the screw *s* so as to balance the centrifugal force. In its place one could also use a string attached to the mass, and threaded through the axis, now considered to be hollow, and loaded at the end more or less heavily by using weights, as indicated in Figure 2.

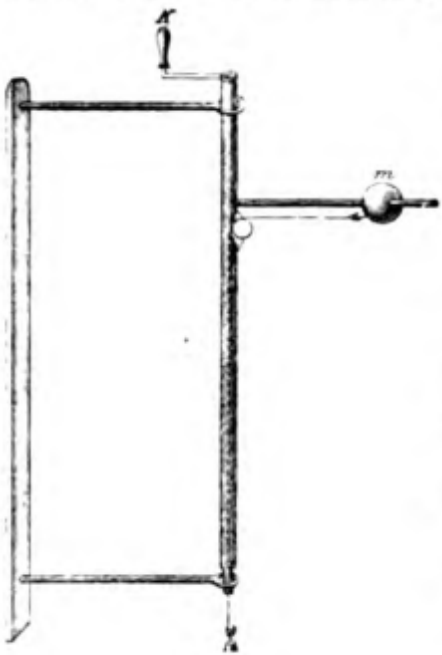

Fig. 2.

In this manner the equilibrium of the mass *m* is adjustable. In the same manner, the axis *CD* supports the mass $\mu$ which is restrained by spring $\Pi$, and adjusted by screw $\sigma$. The distances of the two masses *m* and $\mu$ from their respective axes of rotations are denoted by *r* and $\mu$. The rotational speeds of both axes are related to each other in the following manner. The axis *AB* carries a horizontal gear *E* of radius 1, which engages a vertical gear *F* of the same size; the axis *CD* carries a horizontal friction disk *H* of radius 1, which rubs against a vertical disk G; the latter is attached to the same axis as gear *F*. The friction disc *H* can be fixed at different heights on the axis *CD* using a screw *t*. Thereby the vertical distance *a* between the disk *H* and the axis of the wheels *F* and *G* can be continuously adjusted. If *w* and $\omega$ are the angular velocities of the axes *AB* and *CD*, then the friction wheels ensure that

$$\omega = a w \tag{73}$$

holds. The friction is, of course, an energy consuming force, but if the speed changes are very slow, so that the surfaces rubbing against each other always have infinitesimally different speeds, the energy loss is known to be second order, and also infinitesimal. The connection can therefore be considered as non-dissipating. With the exception of this frictional coupling, all motions should be considered as frictionless, and all components except $m$ and $\mu$ considered as massless. We have here three independent $p_a$ parameters, namely $r$, $\rho$ and $a$, which are varied by the screws $s$, $\sigma$ and $t$, and two $q_b$ parameters, namely $w$ and $\omega$; We select $w$ as von Helmholtz's variable $x$, where $\omega = ax$. The value of $w$ can also be slowly changed from outside using the crank K. Since the forces exerted by the springs $P$ and $\Pi$ are taken to be external then the potential energy due to internal forces is $\Phi=0$. ($\Phi$ can always be set to $0$ if the potential energy depends only on the $p_a$ parameters and not on the $q_b$ parameters, because the former can only be changed by external forces, while the latter do not, by definition, appear in the expression for $\Phi$.) The total kinetic energy is

$$L = -H = \frac{mr^2w^2}{2} + \frac{\mu\rho^2\omega^2}{2}. \tag{74}$$

We set

$$s_1 = -\frac{\partial H}{\partial w} = mr^2 w, \tag{75}$$

$$s_2 = -\frac{\partial H}{\partial \omega} = \mu\rho^2 \omega, \tag{76}$$

so the total energy imparted by the crank K is

$$dQ = w\,ds_1 + \omega\,ds_2 = m\,w\,d(r^2 w) + \mu\,\omega\,d(\rho^2 \omega). \tag{77}$$

Of this the portion

$$mr^2 w\,dw + mw^2 r\,dr + \mu\rho^2 \omega\,d\omega + \mu\omega^2 \rho\,d\rho \tag{78}$$

goes to the increase in kinetic energy.

In addition, the increases in $r$ and $\rho$ perform work

$$mw^2 r\,dr \text{ and } \mu\omega^2 \rho\,d\rho \tag{79}$$

in overcoming the tensions of the springs, which are

$$-\frac{\partial H}{\partial r} = mw^2 r \text{ and } -\frac{\partial H}{\partial \rho} = \mu\omega^2\rho \tag{80}$$

If we introduce $p_a$ and $x$, (that is, in our example $r$, $\rho$, $a$ and $w$) as the independent variables, and we obtain

$$dQ = 2mw^2 r\,dr + 2\mu a^2 w^2 \rho\,d\rho + (mr^2 + \mu a^2 \rho^2)w\,dw \\ + \mu\rho^2 w^2 a\,da, \tag{81}$$

and it is clear that not only is the kinetic energy not an integrating denominator, but that *dQ* has no integrating factor by which it would be transformed into a complete differential of a function of *r, ρ, w* and *a*. This case is furthermore different from the one considered by von Helmholtz insofar as here the constraint introduces a new independent variable *a*. Since, however, *a* is a completely independent quantity, it is easy to devise conditions which make *a* an arbitrary function of $p_a$, that is, of *r,* and *ρ*.

The easiest way is to suppose that the screw *t* is replaced by some kind of direct linkage between the friction disk *H* and the mass *μ*. Assume for example the connection were to be made by a perfectly flexible rod of constant length, the ends of which are fixed to *μ* and *H*. Further assume that one portion of the rod is always parallel to *CD* and the other is always parallel to the transverse arm bearing the mass *μ*. And assume also that the rod, where it bends, runs over a frictionless pulley. Then

$$a = \rho + const. \tag{82}$$

Even easier is to choose the length of the flexible rod as $a = \rho$. This would give

$$dQ = 2mw^2 r dr + 3\mu w^2 \rho^3 d\rho + (mr^2 + \mu\rho^4) w dw \\ = Xdr + Ydr + Zdr, \tag{83}$$

and since

$$X\left(\frac{\partial Y}{\partial w} - \frac{\partial Z}{\partial \rho}\right) + Y\left(\frac{\partial Z}{\partial r} - \frac{\partial X}{\partial w}\right) + Z\left(\frac{\partial X}{\partial \rho} - \frac{\partial Y}{\partial r}\right) = -2m\mu w^3 \rho^3 r \tag{84}$$

the condition for integrability of *dQ* is not met. Of course many other forms for *a* are possible; If instead of the wheel *F* and the disk *G*, any rotating body could be attached to the axis *AB* and connected by a frictionless linkage with the mass *m*; the friction disk *H* would then engage it directly, similar to the apparatus suggested by Helmholtz in his essays. But I have chosen to present the simplest possible example here.

One more consideration must be added here: the rotating masses are not symmetric around the axis of rotation, but it is immediately clear that this is immaterial, and that the same formulas would be applicable to perfectly symmetric rotating bodies with variable moments of inertia. One could indeed transform the system into a Monode by imagining infinitely many axes *AB* and *CD* on which the masses *m* and *μ* occupy all possible angular positions. [11] . However the system shown in Figure 1 is not an Ergode; for if we denote the position angle ∫*wdt* of any of the masses *m* as *W*, and the position angle of any mass *μ* with *Ω*, then over time all possible values of *Ω* will be associated with a certain value of *W*, since the length *a* will in general be irrational. So if the system were an Ergode, then all possible values of pairs of

---

11  Each time a single Monode goes through one of the possible states that could co-exist with it, a single copy can be put in place of the monode. Any changes in magnitudes that are slow, must be uniform enough and slow enough that the changes remain infinitesimal over the time it takes for the system to go through all its states. For such a Monode the designation "isodic" was already suggested earlier.

$$w = \frac{dW}{dt} \quad \text{und} \quad \omega = \frac{d\Omega}{dt} \tag{85}$$

which are compatible with the kinetic energy will occur.

The system illustrated in Figure 1 behaves in this regard as central motion along a non-closed orbit, where certain pairs of values of the rectangular coordinates $x$ and $y$ are incompatible with the kinetic energy. An Ergode would only arise if there were infinitely many such systems side by side, for which the kinetic energy was constant and $a$ had all possible values, so that all possible values of $\omega$ would occur, and $w$ would be determined solely by the kinetic energy equation.

§5

Since the action of the frictional force in the frictional coupling is not obvious, it does not seem superfluous to me to show how the frictional force in the case at hand is equivalent to a set of gears with an infinite number of teeth. Let $O$ be the center and $OP$ the radius of a vertically oriented gear perpendicular to the plane of Figure 3; The teeth are arranged perpendicular to the plane of the gear wheel in many concentric circular rows. Each successive circle has to contain one more tooth than the previous. The distance between two rows of teeth should be infinitesimal, and the thickness of the teeth themselves infinitesimally smaller still, so that a second gear can mesh without any interference with any row of teeth. $QR$ is the radius of this second gear wheel, centered at $O'$, whose plane is horizontal and also perpendicular to the plane of Figure 3.

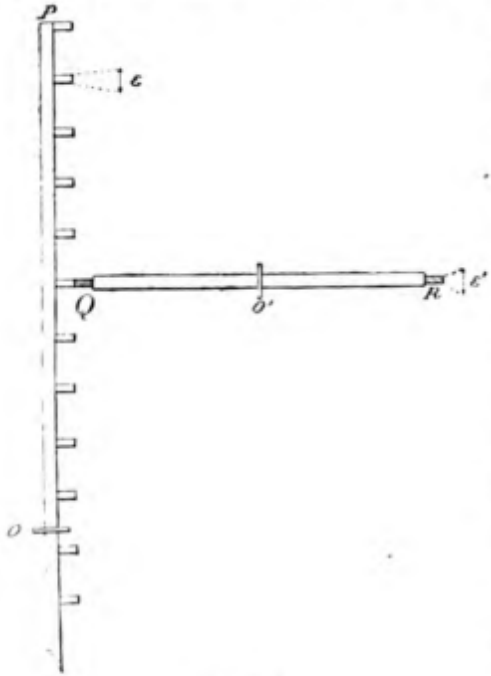

Fig. 3.

This gear has $v$ teeth, which lie in the plane of the gear wheel, and its thickness $\varepsilon'$ is small compared to the distance between the rows of teeth on the first gear. The QR wheel should initially engage with a row of $n-1$ teeth; by shifting parallel to itself, it is lifted out of this row of teeth to smoothly mesh with the next row. It will rotate at its old speed until it meshes with the new set of $n$ teeth; movement beyond this row should be prevented by a suitable stop on the axis. The force between the two gears causes an infinitesimal change in their angular velocity; the original angular velocities are denoted by $w$ and $\omega$, the new ones by $w_1$ and $\omega_1$. The distance between two neighboring teeth, which is the same for both gears, is $\delta$; Then

$$(n-1)w = v\omega, \qquad nw_1 = v\omega_1 \tag{86}$$

$$\frac{(n-1)\delta}{2\pi}, \quad \frac{n\delta}{2\pi} \quad \text{and} \quad \frac{v\delta}{2\pi} \tag{87}$$

will be the radii of the gears, counted from the point of interaction. Let $f$ be the time integral of the total pressure at the point of interaction, then $n\delta f/2\pi$ and $v\delta f/2\pi$ are the moments of the pressure integral with respect to the two axes. $m$ is the angular momentum of the mass associated with the wheel OP from the angular velocity (ie. the moemnt of inertia of the solid body rotating about the axis O, if that mass is rigidly connected tot eh axis of the first wheel); $\mu$ is the equivalent quantity for the second gear wheel. Then

$$2\pi m(w - w_1) = n\delta f, \quad 2\pi \mu(\omega_1 - \omega) = v\delta f, \tag{88}$$

therefore

$$mv(w - w_1) = \mu n(\omega_1 - \omega), \tag{89}$$

from which it follows:

$$w_1 - w = \frac{-w\mu n}{(mv^2 + \mu n^2)}, \tag{90}$$

$$\omega_1 - \omega = \frac{mv^2 \omega}{(n-1)(mv^2 + \mu n^2)}. \tag{91}$$

If the transition from one row of teeth to the other is continuous until $w$, $\omega$ and $n$ have attained the values $W$, $\Omega$ and $N$, then the final result can be obtained by integrating with respect to $n$, using for $dw/dn$ and $d\omega/dn$ the previous equations for $w_1 - w$ and $\omega_1 - \omega$. This yields

$$\frac{dw}{w} = -\frac{\mu n \, dn}{(mv^2 + \mu n^2)}, \tag{92}$$

and by integration

$$W = w\sqrt{\frac{(mv^2 + \mu n^2)}{(mv^2 + \mu N^2)}} \tag{93}$$

while $\Omega$ comes from the equation

$$NW = v\Omega \tag{94}$$

As expected, the energy consumed is infinitesimal, and the assumption made earlier confirms that the total kinetic energy of both wheels remains unchanged by the gradual displacement of the second gear, because the equations

$$mW^2 + \mu\Omega^2 = mw^2 + \mu\omega^2, \quad nw = vw \text{ and } NW = v\Omega \tag{95}$$

generate the same values.

§6

Exactly the same formulas apply to liquid flowing in opposite directions in two channels, if the cross section of the first channel is uniform, but varies in time, and the same is true for the second channel. I want to briefly treat this case here, if only for the purpose of making the very general formulas of von Helmholtz more understandable by the further discussion of one example . Let $r$ be the length, $g$ be the cross section, $s$ the constant density of the liquid in the first channel, and $f$ the fluid velocity. The total liquid mass contained in the first channel, which we wish to denote by analogy with the previous problem by $1/m$, is $sgr$; The mass passing though the cross section in unit time is

$$w = sgf = \frac{f}{rm} \tag{96}$$

The same quantities for the second channel are designated by the corresponding Greek letters. Since there are no other internal forces, then

$$\begin{aligned}\Phi &= 0, \\ L = -H &= \frac{f^2}{2m} + \frac{\phi^2}{2\mu} = \frac{mr^2 w^2}{2} + \frac{\mu\rho^2\omega^2}{2},\end{aligned} \tag{97}$$

and all the previously derived formulas are applicable, only with different meanings of the symbols. The connection between channels can be accomplished by supposing that one immerses in each a paddle wheel whose speed never deviates significantly from that of the liquid due to the resistance of the medium. The rotation rates of both wheels are then coupled as before by friction rollers in a certain ratio which can be changed slowly, as were r and $\rho$. The whole device amounts again to the relation $\omega=aw$ where $a$ is constant with unchanged external conditions, but can, with time, be changed slowly to any possible value. Another generalization is possible here by letting the cross section $g$ be different at different locations. If we designate by $n$ the known liquid mass in one of these cross sections one can then consider $g$ as a slowly varying function of $n$: $\int dn$ is the constant total mass $1/m$ of the fluid. Thereby we can consider $r$ as determined. $dn$ can then be considered as unable of change. Then

$$\begin{aligned}L = -H &= \frac{\int f^2 dn}{2} + \frac{\int \phi^2 dv}{2} \\ &= \frac{w^2}{2s^2}\int \frac{dn}{g^2} + \frac{w^2}{2\sigma^2}\int \frac{dv}{\gamma^2};\end{aligned} \tag{99}$$

The value of $s_b$ for the first channel is

$$\frac{\partial L}{\partial w} = \frac{w}{s^2}\int \frac{dn}{g^2}. \tag{100}$$

The work used to deform the channel is

$$\delta L = -\frac{w^2}{s^2}\int \frac{dn\,\delta g}{g^3}, \tag{101}$$

where $\delta$ is any change produced by deformation of the designated channel. Then

$$dQ = w\,d\left(\frac{w}{s^2}\int\frac{dn}{g^2}\right)+\omega\,d\left(\frac{\omega}{\sigma^2}\int\frac{dv}{\gamma^2}\right) \qquad (102)$$

$$= \frac{w\,dw}{s^2}\int\frac{dn}{g^2} - \frac{2w^2}{s^2}\int\frac{dn\,\delta g}{g^3} + \frac{\omega\,d\omega}{\sigma^2}\int\frac{dv}{\gamma^2} - \frac{2\omega^2}{\sigma^2}\int\frac{dv\,\delta\gamma}{\gamma^3},$$

This expression has again in general no integrating factor when $\omega=wa$, and $w$, $a$, $g$ and $\gamma$ are either independently variable, or $a$ is determined in some way by $g$ and $\gamma$. [12]

*Graz, October 9, 1884.*

**Translators' Notes**
This paper was originally published in German in Crelles Journal (1884-1885) vol. 98, pp. 68-94. We used the text from Ludwig Boltzmann's "Wissenschaftliche Abhandlungen" (Collected Works) published in 1909 by Barth, Leipzig, where it appears as paper number 73 in volume III. Literature citations in the foot notes are given verbatim, except internal references like "Diese Sammlung" (This Collection). To avoid ambiguity these have been changed to Wiss. Abh. Titles of Boltzmann's other papers to which he refers in the text have been translated into English.

Equations were not numbered in the original publication but have been numbered by the translators in order of occurrence.

The word "Kraft" can be variously translated as either force, work or energy depending on context, even though there was a separate German word for energy: Energie.

The term "lebendige Kraft" used by 19[th] c. German physicists would have been translated at the time as 'living force' or 'alive force', but we have used the more modern term 'kinetic energy'.

The specialized words "Monode", "Orthode", "Ergode", "Holode", "Planode" and "isodic", some coined by Boltzmann in this and earlier work, are not easy to translate and thus have been left as is. Their meanings must be determined by the definitions and equations provided by Boltzmann. Nevertheless, concepts designated by these terms are clearly fore-runners of the ensembles - micro-canonical and canonical - defined later by Gibbs in 1902 in his "Elementary Principles in Statistical Mechanics". Ergode also gave its name to the concepts of ergodic systems and ergodicity which have played an important, if contentious, part in the foundations of statistical mechanics.

Kim Sharp, University of Pennsylvania.  sharpk@upenn.edu
Franz Matschinsky, University of Pennsylvania

---

12 Immediately before the dispatch of the last proof sheet I saw in the previous issue of the Crelle Journals the 2[nd] essay of Herr von Helmholtz on this subject, clearer, and apparently just as important and content-rich as the first one. Of course, I am no longer able to use it in any way here, and I only notice that my objections to the deductions given by Helmholtz in §8 seem to be to be tenable, if the same should prove that in all physically possible monocyclic systems the kinetic energy is unconditionally an integrating denominator of *dQ*. Rather, this deduction seems merely to show me that if any integrating factors exist, one of them must be the reciprocal of the kinetic energy. Graz, December 8, 1884.